\documentclass[preprint,3p]{elsarticle}

\usepackage{amsmath,amssymb,amsfonts}
\usepackage{graphicx}
\usepackage[hidelinks]{hyperref}
\usepackage{float}
\usepackage{svg}
\usepackage{tabularx}
\usepackage{booktabs}
\usepackage{multirow}
\usepackage{pifont}
\usepackage{threeparttable}
\usepackage[linesnumbered, lined, ruled, vlined]{algorithm2e}

\newcommand{\cmark}{\ding{51}}%
\newcommand{\xmark}{\ding{55}}%

\begin{document}
\title{FunnelNet: An End-to-End Deep Learning Framework to Monitor Digital Heart Murmur in Real-Time}

\author[1]{Md Jobayer}
\ead{mdjob331@student.liu.se}

\author[2]{Md Mehedi Hasan Shawon}
\ead{mshawon@umd.edu}

\author[3]{Md Zakir Hossain}
\ead{zakir.hossain@anu.edu.au}

\author[4,5]{Shreya Ghosh}
\ead{shreya.ghosh@uq.edu.au}

\author[6]{Imre Rudas}
\ead{rudas@uni-obuda.hu}

\author[3,4,6]{Tom~Gedeon}
\ead{tom.gedeon@curtin.edu.au}

\author[4]{Md Rakibul Hasan\corref{cor1}}
\ead{rakibul.hasan@curtin.edu.au}

\cortext[cor1]{Corresponding author}


\affiliation[1]{
organization={Department of Biomedical Engineering, Linköping University},
city={Linköping 58183},
country={Sweden}
}

\affiliation[2]{
organization={Department of Electrical and Computer Engineering, University of Maryland College Park},
city={Maryland 20742},
country={USA}
}

\affiliation[3]{
organization={School of Medicine and Psychology, The Australian National University},
city={Canberra},
state={ACT 2601},
country={Australia}
}

\affiliation[4]{
organization={School of Electrical Engineering, Computing and Mathematical Sciences, Curtin University},
city={Bentley},
state={WA 6102},
country={Australia}
}

\affiliation[5]{
organization={School of Electrical Engineering and Computer Science, The University of Queensland},
city={Brisbane},
state={QLD 4072},
country={Australia}
}

\affiliation[6]{
organization={University of ÓBuda},
city={Budapest},
country={Hungary}
}

\begin{abstract}

Heart murmurs are abnormal sounds caused by turbulent blood flow in the heart. Several diagnostic methods are available to detect heart murmurs and their severity, including cardiac auscultation, echocardiography, and phonocardiography (PCG). However, these methods have limitations, including the need for extensive training among healthcare providers, the cost and accessibility of echocardiography, and noise interference during PCG data processing. This study proposes an end-to-end real-time heart murmur detection approach using traditional and depthwise separable convolutional networks. We applied a Butterworth filter and Continuous Wavelet Transform (CWT) to eliminate noise and extract meaningful features from the PCG data. The proposed network consists of three parts: a Squeeze net that generates a compressed data representation, a Bottleneck layer that minimizes computational complexity using depthwise-separable convolutions, and an Expansion net that up-samples the data to capture fine details. We evaluated our model on the publicly available CirCor pediatric heart sound dataset. Using only $\sim$5.4k parameters, we achieved an accuracy of 85\%, a sensitivity of 85\%, and a specificity of 92\%, successfully outperforming several larger models. Furthermore, we converted our network into a TinyML format and tested it on two resource-constrained devices, achieving an average real-time inference accuracy of 91\% on a Raspberry Pi 4B and 80\% on an Android smartphone. The proposed lightweight model offers a robust deep learning framework for accurate, real-time heart murmur detection, showing strong promise for accessible medical diagnostics in limited-resource environments. The code is publicly available at \url{https://github.com/jobayer/FunnelNet}.

\end{abstract}


\maketitle

\section{Introduction}

\begin{figure}[htp]
    \centering
    \includegraphics[width=0.75\textwidth]{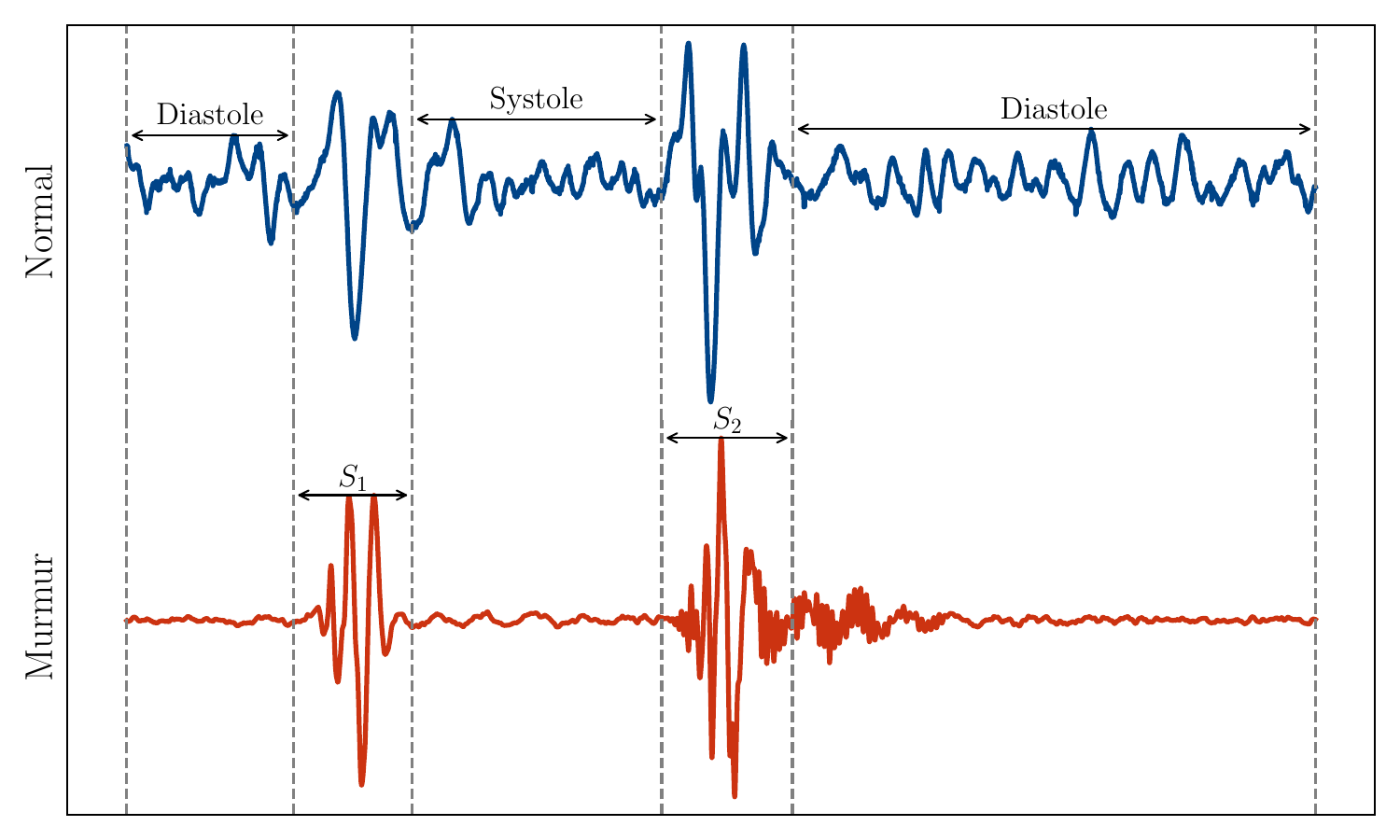}
    \caption{Heart's cyclic pattern. It consists of two main phases: systole and diastole. S1 and S2 are the two main heart sounds at the beginning of systole and diastole, respectively.}
    \label{fig_heart_patterns}
\end{figure}

The cardiac cycle is the sequence of pressure changes within the heart, which leads to blood movement in different heart chambers and throughout the body \cite{pollock_physiology_2024}. Among other states, as we can see in \autoref{fig_heart_patterns}, four leading states of interest are systole, diastole, S1, and S2. Here, systole is muscle contraction, diastole is the relaxation phase, and S1 and S2 are the first and second heart sounds. A heart murmur is an abnormal sound heard during the cardiac cycle, typically due to turbulent blood flow within the heart. Heart murmurs can be mild or severe, indicating underlying heart disease. Among several traditional techniques, cardiac auscultation, which involves listening to the heart with a stethoscope, is still a cost-effective screening method and provides valuable information about the mechanical activities of the heart \cite{oliveira_circor_2022}. However, auscultation is a difficult skill to master, which requires extensive training accompanied by continuous clinical experience \cite{lim_ai_2021}. Echocardiography is another diagnostic tool to visualize the heart structures and assess abnormalities associated with a murmur. It is particularly important for patients with systolic murmurs of unknown causes suspected of having significant heart disease \cite{attenhofer_jost_echocardiography_2000}. It also takes considerable time to conduct diagnostic procedures and is not easily accessible \cite{draper_murmur_2019}. Furthermore, traditional methods provide limited quantitative measurements of heart murmurs, which can lead to misdiagnosed cases. In general, reliance on manual auscultation can be time-consuming, which may not always be feasible in a busy clinical setting, highlighting the need for more efficient and standardized approaches to cardiac evaluation \citep{reyna_heart_2022}.

Computational methods, including machine learning algorithms, are becoming promising tools offering more consistent results, reducing the dependence on human interpretation. The limitations of conventional approaches for detecting heart murmurs can also be overcome by automated analysis of heart sound recordings utilizing deep learning frameworks, offering a more standardized and accessible means  \citep{chorba_deep_2021}. However, some of these approaches are resource-intensive and require special hardware requirements. Therefore, a lightweight machine learning algorithm would be a more efficient and practical solution. In particular, they can provide a scalable solution for assessing medical conditions in remote health monitoring applications by simplifying computational complexity while maintaining high performance.

The primary contributions of the paper are as follows:
\begin{itemize}
    \item We proposed an end-to-end approach for heart murmur detection using a separable convolutional neural network (CNN) model.
    \item We demonstrated the real-time detection capability of our model using two resource-constrained devices.
    \item Our proposed model achieved comparable performance on CirCor dataset in several performance metrics.
    \item The TinyML version of our model outperforms some of the regular SOTA models.
\end{itemize}

\section{Related Work}

\subsection{Heart Murmur Detection}

The detection of heart murmurs using machine learning has gained significant attention in recent years due to its potential to improve the diagnosis of cardiovascular diseases. Researchers have explored different approaches to classify heart sounds and detect abnormalities accurately. \citet{keikhosrokiani_heartbeat_2023} proposed a hybrid adaptive neuro-fuzzy inference system (ANFIS) and an artificial bee colony approach for classifying heartbeat sounds. This study used Mel frequency cepstral coefficients feature extraction methods to detect and classify abnormal heart sounds. However, ANFIS generates complex models and is computationally expensive even for simpler problems \cite{senthilselvi_performance_2021}. Similarly, \citet{raza_heartbeat_2019} proposed a Recurrent Neural Network model for classifying heartbeat sound signals. In this paper, they used down-sampling techniques to extract discriminant features and reduced the dimension of the frame for lower computational power. However, they had to sacrifice the model performance, yielding a maximum accuracy of 80.8\%. Moreover, \citet{shuvo_cardioxnet:_2021} developed CardioXNet, a streamlined deep learning model for categorizing cardiovascular disorders using heart sound data. This study showed the importance of S1 and S2 sounds and irregular variants such as S3, S4, and murmurs in pathological inference. Although they achieved satisfactory results in accuracy compared to other SOTA models, they still lack performance robustness and lower resource consumption. Furthermore, \citet{xiao_heart_2020} proposed a 1-D CNN with low parameter consumption for heart sound classification, demonstrating the effectiveness of the memory-efficient models. Nevertheless, they lack automatic environmental noise suppression techniques, making them less reliable. \citet{cheng_heart_2023} introduced a novel heart sound classification network based on convolution and transformer while \citet{ghosh_automated_2020} automated the detection of heart valve diseases using a multiclass composite classifier with PCG signals. Furthermore, \citet{chen_classifying_2023} analyzed the MelSpectrum and Log-MelSpectrum characteristics of heart sound signals, demonstrating the importance of feature selection in improving classification accuracy. In summary, researchers have used different machine learning models, including deep learning architectures, CNNs, and ensemble methods, to detect heart murmurs effectively. However, none of them could maintain good performance and low memory consumption, providing a real-time and robust solution. \autoref{tab_dfsf_overview} represents an overview of the existing literature with their signal properties and methodologies.

\begin{table}[ht]
\centering
\caption{A brief overview of used datasets, types of extracted features, the audio segmentation, and the filters to denoise audio data from the existing literature. The filters are expressed as ${\text{[Filter name]}}_{\text{Frequency in Hertz}}^{\text{Order}}$. BP, LP, and HP mean band-pass, low-pass, and high-pass, respectively.}
\label{tab_dfsf_overview}
\begin{tabularx}{\textwidth}{@{}XXXX@{}}
\toprule
Reference & Extracted Features & Segmentation & Filters \\ \midrule
\citet{li_heart_2021} & None & 2s & ${\text{[Butterworth BP]}}_{\text{25-500}}^{\text{5}}$ \\
\citet{cheng_heart_2023} & None & 2.5s & ${\text{[Butterworth BP]}}_{\text{25-400}}^{\text{4}}$ \\
\citet{kui_heart_2021} & Log MFSC & Dynamic & None \\
\citet{li_heart_2022} & MFCC & 2s & ${\text{[Butterworth BP]}}_{\text{400}}^{\text{5}}$ \\
\citet{shuvo_cardioxnet:_2021} & None & 1.125s & None \\
\citet{krishnan_automated_2020} & None & 6s & ${\text{[Savitzky-Golay]}}_{\text{500}}$ \\
\citet{noman_short-segment_2019} & MFCC (CNN2D) & Not mentioned & ${\text{[Butterworth BP]}}_{\text{25-400}}$ \\
\citet{chen_heart_2023} & CWT & 1.6s & ${\text{[LP]}}_{\text{1000}}$ \\
\citet{chowdhury_time-frequency_2020} & DWT, Mel-scaled PSD and MFCC & Not mentioned & [LP] and [HP] \\
\citet{ghosh_automated_2020} & Chirplet transform (CT) & Not mentioned & ${\text{[Butterworth BP]}}_{\text{25-900}}$ \\
FunnelNet (ours) & Morlet-CWT & 5s & ${\text{[Butterworth BP]}}_{\text{20-500}}^{\text{2}}$ \\ \bottomrule
\end{tabularx}
\end{table}

\subsection{Tiny Machine Learning (TinyML)}

TinyML is an emerging field that develops machine learning algorithms that can run on resource-constrained devices such as  Internet of Things (IoT) units, edge devices, and embedded systems. In recent years, there has been a growing interest in applying TinyML technology in the healthcare sector, with applications ranging from ECG monitoring to posture detection to wearable healthcare \citep{kim_tinyml-based_2023, dr_g_sophia_reena_posture_2023, sabry_machine_2022}. \citet{kim_tinyml-based_2023} presents a study on TinyML-based classification in an embedded ECG monitoring system. Their work showcases the application of CNNs in efficiently processing ECG data. Furthermore, \citet{dr_g_sophia_reena_posture_2023} introduces a posture guardian system that uses TinyML for smart muscle strain detection and correction. This application demonstrates how TinyML can assess and categorize real-time data, providing immediate feedback and corrective measures. Such innovations promise to improve patient outcomes and prevent musculoskeletal injuries through proactive monitoring and intervention. Furthermore, \citet{sun_case_2023} advocates adopting TinyML in healthcare by proposing CNNs to estimate blood pressure on the real-time edge. Highlights the feasibility of using TinyML for critical healthcare tasks without compromising data privacy or security. In addition, \citet{sabry_machine_2022} discusses the more comprehensive scope of machine learning applications in healthcare wearable devices, highlighting the importance of exploring TinyML-embedded solutions and optimization techniques in the IoT ecosystem. Regarding heart sound detection, lightweight end-to-end neural network models have been designed specifically for automatic heart sound classification, demonstrating the feasibility of using such models in edge devices \citep{li_lightweight_2021}. Overall, applying TinyML in the medical domain represents a significant advance in healthcare technology, such as real-time monitoring, personalized care, and improved diagnostic accuracy.

\section{Methods}

\subsection{Wavelet Transformation}

\subsubsection{Wavelet}

A wavelet is a mathematical function that provides a localized representation of signals in both time and frequency domains confined in a finite interval within a specific domain of interest. Unlike the traditional Fourier transformation technique, which is mainly useful in non-stationary signals, wavelet functions are very good at representing a signal's localized features and transient events. Mathematically, this function can be expressed as follows.

\begin{equation}\label{eq_wavlet}
    \psi_{a, b}(t) = \dfrac{1}{\sqrt{\mid a \mid}} \psi (\dfrac{t-b}{a})
\end{equation}

This equation describes how a signal is dilated and shifted by the parameters $a$ and $b$. Here $a$ is the scaling parameter that determines the width of the signal and $b$ is the translating parameter that shifts the signal on the time axis. The $\psi (t)$ is the mother wavelet function, which acts as a bandpass filter. The factor $1/\sqrt{\mid a \mid}$ is called the energy-preserving factor, which ensures that the energy of the wavelet function remains conserved on different scales.

\subsubsection{Morlet Wavelet}

Among many other mother wavelets $\psi (t)$, the Morlet wavelet is a complex-valued function that exhibits a Gaussian envelope modulated by a sinusoidal oscillation, shown as follows:

\begin{equation}\label{eq_morlet}
    \psi (t) = A \cdot e^{- \left(\dfrac{t^2}{2 \sigma^2}\right)} \cdot e^{iwt}
\end{equation}

Like the wavelet function, $A$ is a normalization factor that ensures that the energy distribution among the axes is conserved. In addition, we have two other main components in the wavelet function: a Gaussian window $\exp{[- \left(t^2/2 \sigma^2\right)]}$ and a complex sine function $\exp{[iwt]}$ where $w = 2 \pi f$. Here, $i$ is the imaginary operator $(i=\sqrt{-1})$ making the sine wave complex, $f$ is the frequency in Hertz, and $t$ is the time in seconds. The $\sigma=n / 2 \pi f$ is the standard deviation that controls the width of the Gaussian envelope, and $n$ is the number of cycles \citep{cohen_better_2019}. The Morlet wavelet is an ideal mother wavelet for applications such as heart sound because it is well suited for the oscillatory components of such signals \citep{tiscareno_review_2018}.

\subsubsection{Continuous Wavelet Transform}

Continuous Wavelet Transform (CWT) \citep{goupillaud_cycle_octave_1984} is a decomposition technique that works for stationary and non-stationary signals that can represent an analog signal $x(t)$ based on two varying parameters: the time shift parameter $b$ and the scale parameter $a$.

\begin{equation}\label{eq_cwt}
    CWT\{ x(t); a, b \} = \int x(t) \; \psi_{a, b}^{*}(t) \; d(t)
\end{equation}

Compared to the conventional short-time Fourier transform, which uses the same window length, CWT uses short and long windows at high and low frequencies, respectively. This helps to obtain a better resolution from the CWT decomposition compared to the other techniques \citep{taebi_time_frequency_2017}. Apart from that, compared to other decomposition methods like Empirical Mode Decomposition and Variational Model Decomposition, CWT has high frequency-time resolution \citep{okpok_adaptive_2024} and less sensitivity to noise due to issues like modal mixing and false modes due to over- and under-enveloping \citep{han_vibration_2018}. The Morlet wavelet, in conjunction with CWT, provides the most accurate instantaneous frequencies for the time-frequency distribution signals compared to other mother wavelet-based CWT \citep{taebi_time_frequency_2017}. In heart sound applications, most frequencies range from 20--600 Hz \citep{duan_bionic_2021}, making CWT an effective tool for decomposing low-frequency signals like heart sound, offering a dynamic window length.

\subsection{Outlier Removal}

\begin{figure}[htp]
    \centering
    \includegraphics[width=0.9\textwidth]{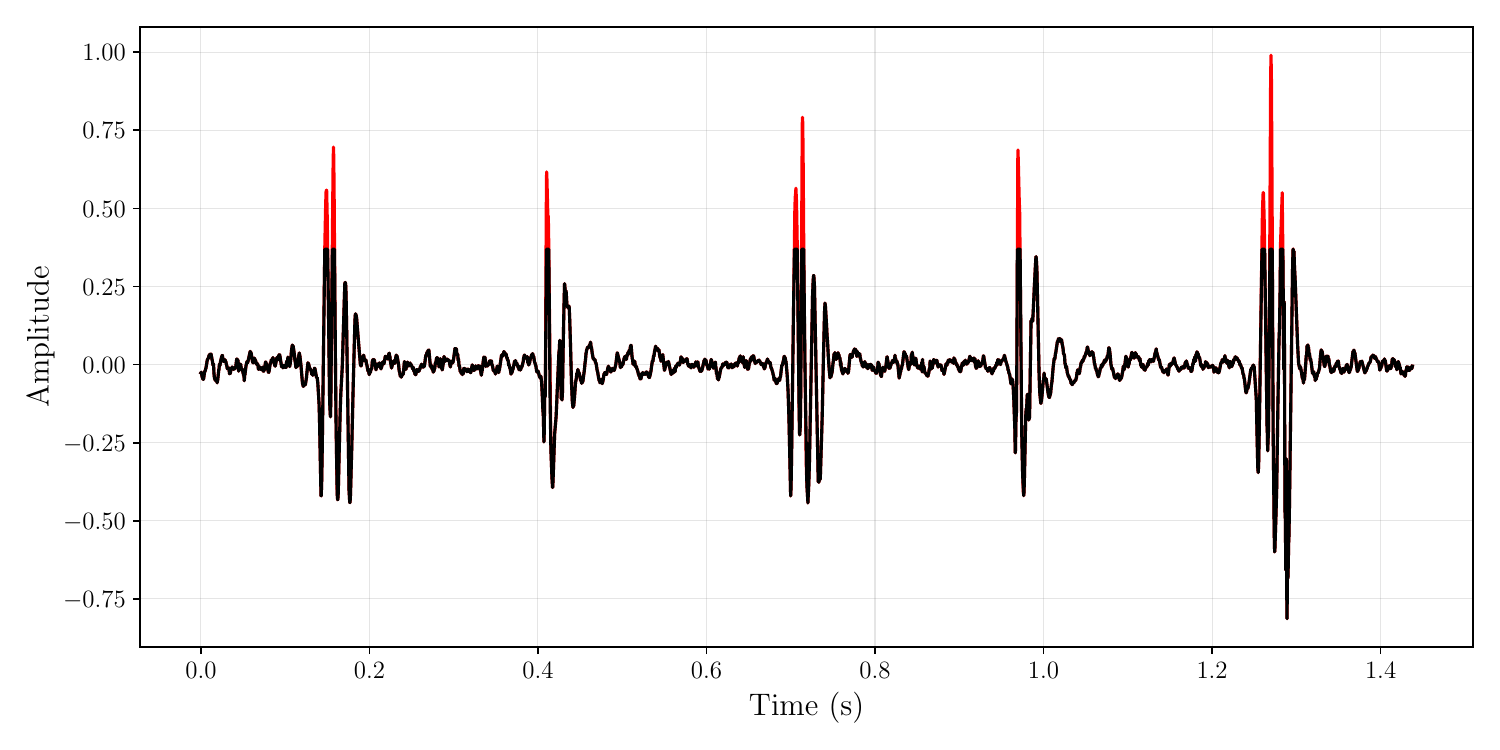}
    \caption{A sample heart sound audio file without any preprocessing steps applied to it. The black color line plot is the final plot after removing the outliers, where the red-colored line plot is the outlier and is removed using the mean and standard deviation relationship of the audio data as shown in \autoref{eq_outlier}.}
    \label{fig_audio_n_outliers}
\end{figure}

In the audio dataset, particularly for heart sounds, most of the sample data of all classes belong to a particular frequency, and only a few have large spikes, usually unwanted noises. So, we provided a conditional equation to remove these spikes or outliers based on their distribution in the time domain.

Let $x$ be the sequence of audio data points represented by $x = \{ x_1, x_2, x_3 \ldots x_n \}$ and $\mu$ be the mean of the sequence of audio data points.

Also, $\sigma$ is the standard deviation of the audio data points:
\begin{equation}\label{eq_outlier}
\begin{aligned}
\sigma &= \sqrt{\dfrac{1}{n} \sum_{i=1}^{n} (x_i - \mu)^2} \\
t &= \mu + p \times \sigma \\
\Tilde{x} (x, p) &=  \begin{cases} 
x_i & \text{if } x_i \leq t \\
t & \text{if } x_i > t
\end{cases}
\end{aligned}
\end{equation}

Here, $t$ is the conditional limit for the original data point $x_i \in x$, and $p$ is the multiplicative factor known as \emph{patience}, which determines the strictness of the acceptance rate of the input amplitudes. In our experiment, we used $p=3$ as the \emph{patience} value to determine an input as an outlier. If $x$ crosses the limit in amplitude compared to $t$, then $x$ is cut to a maximum value of $t$, producing the outlier-free signal $\Tilde{x}$ as shown in \autoref{fig_audio_n_outliers}.

\subsection{Noise Removal}

Although most spikes are omitted after removing the outliers, the remaining signals still have significant magnitudes. This might explain the increased computational power requirement, which is crucial in resource-constrained devices running on a limited energy source. As discussed earlier, heart sounds usually have a very low and specific frequency range; our paper selected a Butterworth band-pass filter to allow only those particular frequencies. While CWT effectively removes preliminary noise levels, especially those that do not align with the frequency characteristics of interest, it is insufficient on its own to eliminate persistent noise that overlaps with the signal of interest. Using a Butterworth filter in the signal-processing domain effectively isolates the relevant frequency components associated with the input signals while suppressing unwanted noise and artifacts. It balances filtering efficiency and frequency response characteristics well for isolating phase-sensitive signals from audio data.

\begin{figure}[htp]
    \centering
    \includegraphics[width=0.75\textwidth]{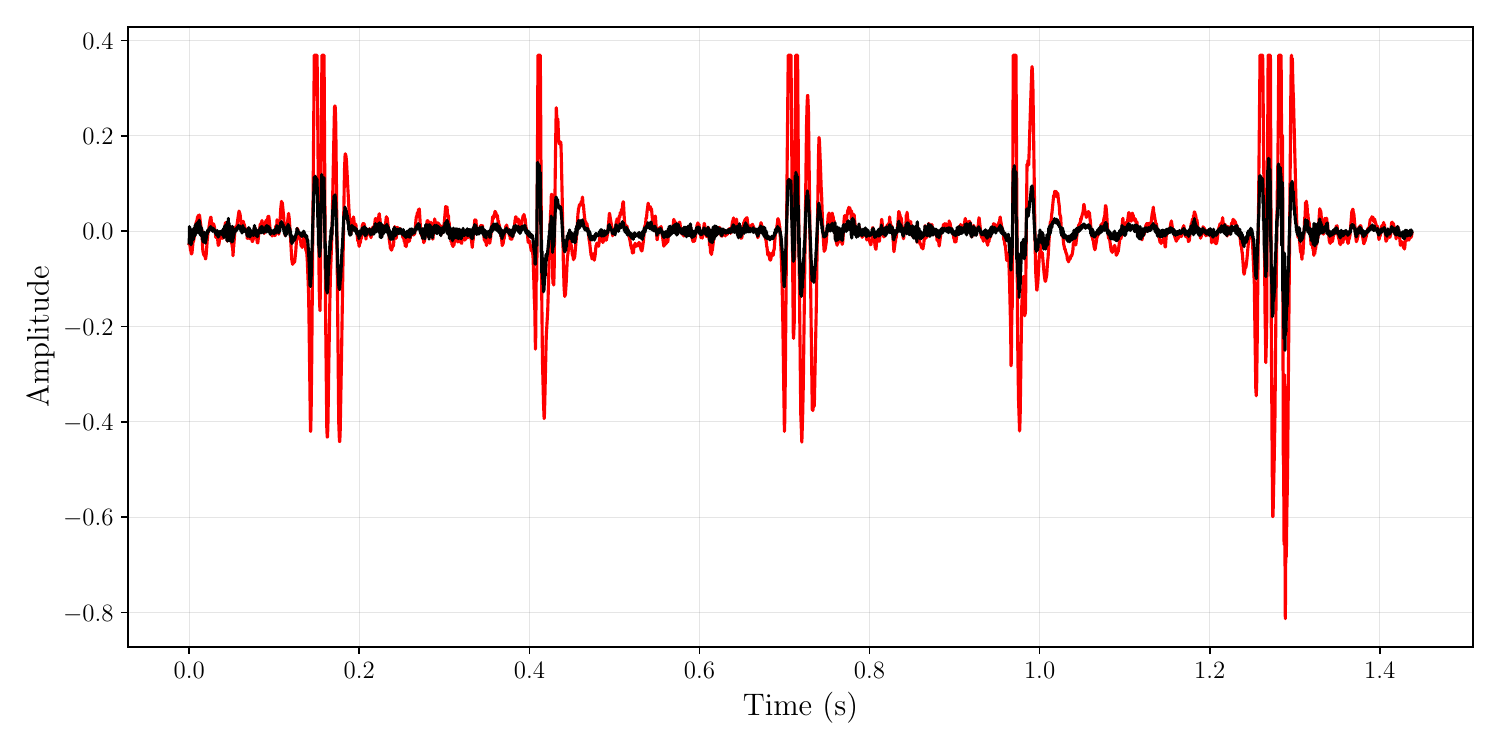}
    \caption{A sample heart sound audio file after removing the outliers from it. The black-colored line plot is the plot after the Butterworth band-pass filter is applied to it. The red line plot is the discarded plot filtered by the Butterworth filter.}
    \label{fig_audio_n_bwf}
\end{figure}

A Butterworth band-pass filter is a signal-processing filter allowing certain ranges of frequencies while attenuating frequencies outside this range. The Butterworth filter is characterized by a maximally flat frequency response in the passband and a gradual roll-off in the stopband. This filter is known to be advantageous for its simplicity, stability, and smooth frequency response. Although a higher-order filter does a better job of filtering the desired frequencies \citep{shelishiyah_signal_2022}, a larger number could also be a reason for information loss \citep{chen_optical_2013}, especially in the small frequency range. Therefore, in our paper, we have used a second-order Butterworth filter with a high cut of $f_H = 500$ Hz and a low cut of $f_L = 20$ Hz to capture all kinds of information from the signal.

\begin{figure}[htb]
    \centering
    \includegraphics[width=0.6\textwidth]{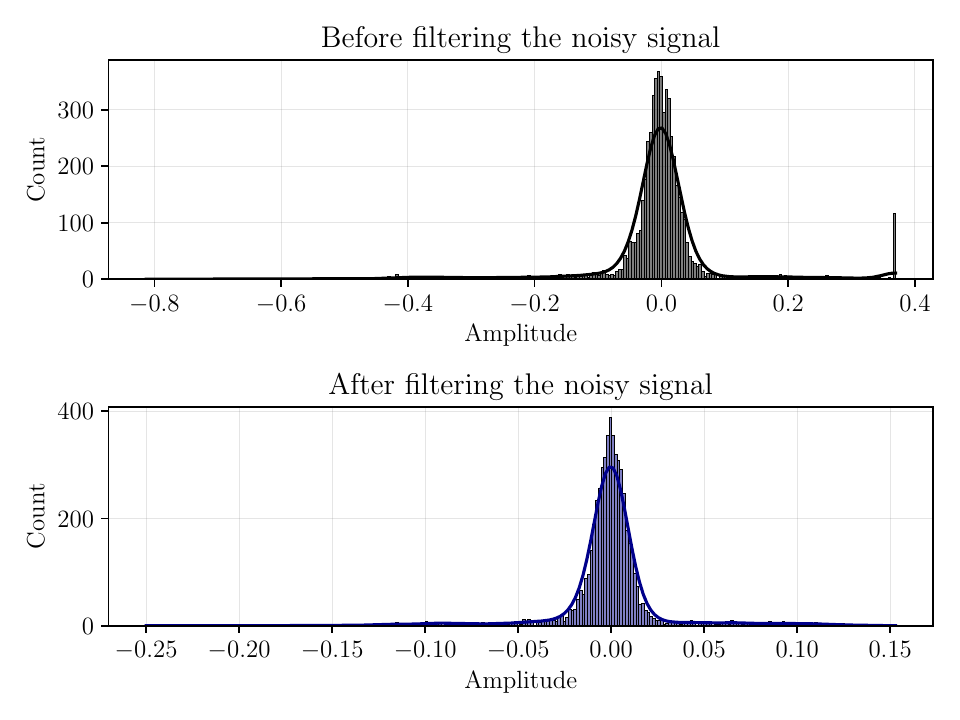}
    \caption{Distribution plots for a sample audio file before and after the Butterworth filter has been applied show that their densities changed from a certain range to a narrower one.}
    \label{fig_audio_n_bwf_hist}
\end{figure}

As we can see in \autoref{fig_audio_n_bwf}, most of their amplitude varied between $[-0.4, 0.4]$ before passing the signals through the Butterworth filter. As soon as it passed through the filter, most frequencies were reduced to a range of $[-0.2, 0.2]$. In addition, it is noticeable from \autoref{fig_audio_n_bwf_hist} that the data was scattered more sparsely before being filtered with a Butterworth filter. After applying the filter, most of the noise from the data was filtered out, and now they are located closer to each other.

\subsection{Data Oversampling}

\begin{figure}[!tb]
    \centering
    \includegraphics[width=0.4\textwidth]{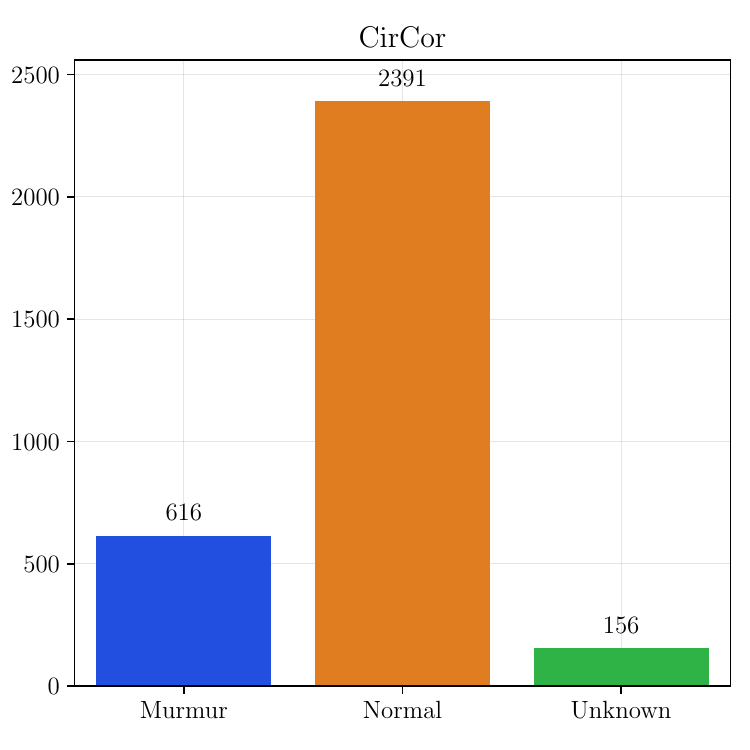}
    \caption{Class distribution of the CirCor dataset which is heavily biased towards the `Normal' class.}
    \label{fig_cls_dist}
\end{figure}

It is often found that the data are quite imbalanced and heavily biased toward a particular class in a dataset. This scenario usually overfits a network and carries weight toward a particular class, which is undesirable. Two main options to overcome this issue are to down-sample or over-sample our data. If the number of samples in the rest of the classes is negligible, the network will not have enough data to learn from if we down-sample the data. Hence, it is better to choose the oversampling technique to create synthetic data that mimics the original ones, creating an equal number of samples for each class. The Synthetic Minority Oversampling Technique (SMOTE) \citep{chawla_smote_2011} is a popular method to address this problem for unbalanced datasets. This method produces synthetic instances for the underrepresented class to equalize the distribution of classes. SMOTE creates new synthetic instances by interpolating between existing minority-class instances. The algorithm selects a minority class instance and then finds its k nearest neighbors in the feature space depending on the oversampling requirement. A new instance is then generated by randomly selecting one of these neighbors and creating a synthetic example along the line connecting the selected neighbor and the original minority class instance. The class distributions of the CirCor dataset is illustrated in  \autoref{fig_cls_dist}.

\subsection{Model Architecture}

\begin{figure*}[htb]
    \centering
    \includegraphics[width=\textwidth]{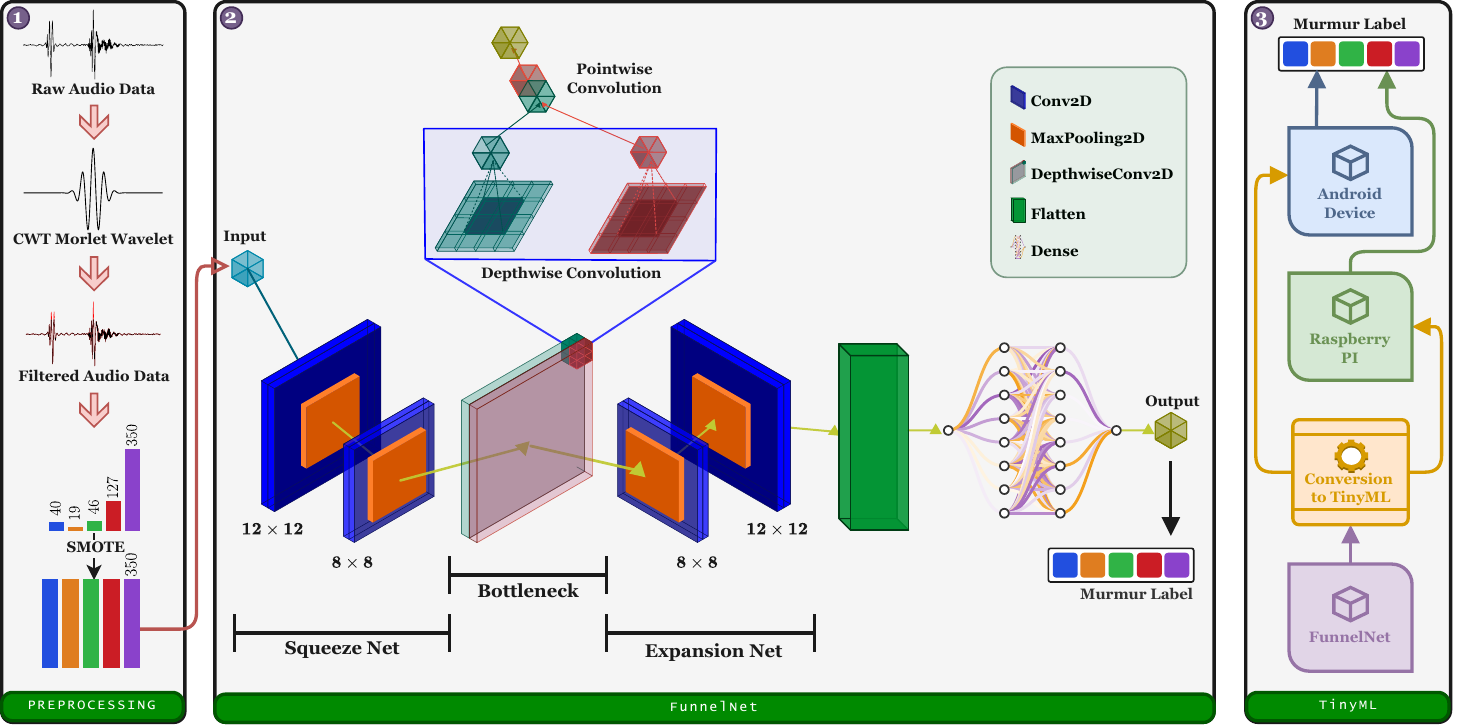}
    \caption{An illustration of the proposed FunnelNet network architecture, divided into three parts: the squeeze net, the bottleneck, and the expansion net. The Squeeze and Expansion networks work like an encoder and decoder to compress and upscale the most relevant input features. A depthwise CNN has been employed at the Bottleneck for reduced computational complexity. A fully connected layer is placed at the end to predict the correct class.}
    \label{fig_proposed_model}
\end{figure*}

Our proposed FunnelNet network, as shown in \autoref{fig_proposed_model}, was designed to prioritize lightweights while delivering competitive results compared to other SOTA models. The name `FunnelNet' derives from its architectural shape, which resembles two funnels facing each other and connected. The network's layers can be divided into conventional CNN, depthwise and pointwise convolutional layers, and fully connected (FC) layers. These can then be summarized into three parts: the Squeeze net, the Bottleneck, and the Expansion net. We got the motivation behind this configuration from the UNet \citep{ronneberger_unet_2015} architecture, which is known for its superior performance in the case of medical imaging and other domains in general. However, we reduced the total number of layers and their order to maintain the lightweight. We replaced the bottleneck with layers of CNNs, depthwise and pointwise, motivated by the MobileNet architecture \citep{howard_mobilenets_2017}, which is known to be very fast and efficient in computational complexity.

\begin{algorithm}[!htb]
    \DontPrintSemicolon
    \caption{End-to-end FunnelNet Architecture}\label{algo:arch}
    \KwIn{RAW audio data in \emph{.wav} format}
    \KwOut{Binary or multiclass prediction}
    \KwData{Audio file paths as a list}

    $\mathcal{A} \gets \emptyset$ \tcp*[h]{empty list} \;
    
    \ForEach{$path$ \textbf{in} $\exists paths$} {
		$file \gets readAudioData(path)$\;
		$file \gets \dfrac{file}{\exists quantFactor}$ \tcp*[f]{quantize 16-bit PCM WAV file}\;
		\If{$\exists freq$ \textbf{is} $\neg None$}{
			$file \gets resample(file, \exists freq)$\;
		}
		$\mathcal{A} \gets \mathcal{A} + file$\;
    }
    
    $\mathcal{A} \gets removeOutliers(\exists patience, \mathcal{A})$ \tcp*[f]{see \autoref{eq_outlier}} \;

    $\mathcal{A} \gets removeNoise(\mathcal{A})$ \tcp*[f]{using Butterworth} \;
    
    $\mathcal{C} \gets \emptyset$ \;
    
    \ForEach{$item$ \textbf{in} $\mathcal{A}$}{
    	$cwt \gets computeCwt(item)$ \tcp*[f]{see \autoref{eq_cwt}} \;
    	$cwt \gets T(cwt)$ \tcp*[f]{transpose the signal}\;
    	$cwt \gets \mathbb{R}(cwt)$\;
    	$\mathcal{C} \gets \mathcal{C} + cwt$
    }
    
    $\mathcal{X} \gets normalize(\mathcal{C})$ \;
    $\mathcal{Y}_{X.shape[0]} \gets \mathcal{U} \in \mathbb{N}$ \tcp*[f]{murmur label} \;
    $\hat{\mathcal{X}}, \hat{\mathcal{Y}} \gets smote(\mathcal{X}, \mathcal{Y})$ \;
    $folds \gets skf(\exists k, \hat{\mathcal{X}}, \hat{\mathcal{Y}})$ \tcp*[f]{stratified $k$-fold}\;
    $\mathcal{M} \gets FunnelNet()$ \;
    $modelTrain(\mathcal{M}, folds, \ldots, \hat{\mathcal{X}}, \hat{\mathcal{Y}})$\;
    $output \gets \mathcal{M}(\hat{\mathcal{X}})$ \tcp*[f]{model's prediction} \;

\end{algorithm}

\subsubsection{Squeeze Net}

The squeeze phase of the network works like an encoder and consists of two CNN layers with filter sizes of 16 and 8, respectively. In both layers, a kernel size of $2 \times 2$ is used. The activation function utilized in these layers is `tanh' due to its $[-1, 1]$ output range, which suits our signal data containing values of the same range. This Squeeze net aims to extract the most relevant information while discarding redundant ones. It compresses the input into a latent space representation that contains the most critical features. This reduction in spatial dimensions helps capture high-level abstract representations of the input data.

\subsubsection{Bottleneck}

Following the squeezing phase, we employ a depthwise separable CNN known for its computational efficiency compared to traditional approaches. This involves two main components: a depthwise convolution that uses filters with a depth of 1 for each input channel and a pointwise convolution with $1 \times 1$ filters to consolidate information. Like the squeeze net, we have a kernel size of $2 \times 2$ with an activation function of `tanh. Compared with conventional CNN, it significantly reduces parameter count, mitigating overfitting and resulting in smaller model sizes, which is desirable for TinyML models.

\subsubsection{Expansion Net}

The subsequent expanding phase, which works like a decoder, mirrors the squeezing phase but in reverse order, utilizing conventional layers with the same filter and kernel sizes. As a decoder network, the expansion net aims to reconstruct the original input from the compressed representation generated by the encoder or the squeeze net. This way, it tries to recover the spatial details to produce meaningful outputs. 

After each convolutional layer, including the depthwise separable convolution layer, we apply a maximum pooling layer with a pool size of $2 \times 2$. This is succeeded by a Flatten layer to linearly represent the data, followed by an FC layer with a depth of 8 and using the ReLU activation function. Finally, we incorporated a fully connected layer with a depth of three and applied a Softmax activation function to classify the three categories of the CirCor dataset. Algorithm \ref{algo:arch} explains the workflow of our proposed solution for the automated heart murmur prediction mechanism.

\begin{table}[!htb]
\centering
\caption{Bayesian optimization-based Hyperparameter Tuning}
\label{tab_hp}
\begin{tabular}{@{}lll@{}}
\toprule
Parameter & Range & Best Value \\ \midrule
Optimizer & [adam, adamw, sgd, rmsprop] & adam \\
Learning Rate & [0.0001:0.01] & 0.005 \\
Batch Size & [32:32:128] & 32 \\ \bottomrule
\end{tabular}
\end{table}

\section{Experiments}

\subsection{Dataset}\label{sec_ds_desc}




We chose the CirCor DigiScope Phonocardiogram \citep{oliveira_circor_2022} dataset to evaluate the proposed model. It is a large publicly accessible pediatric heart sound dataset with 5,272 recordings. This dataset contains data from 1,568 patients that span an age range of 0 to 21 years. There are a total of 33.5 hours of recordings, where the duration of the individual recordings ranges from 4.8 to 80.4 seconds. The dataset contains three classes: `normal,' `murmur,' and `unknown.' The `unknown' class was unidentifiable by the annotator as it did not meet the required signal quality standards. The dataset also contains external noises that pose realistic diagnostic challenges. For convenience, we refer to `CirCor' as the alias of this dataset throughout the paper.

\subsection{Hyperparameter Tuning}

We conducted hyperparameter tuning using Bayesian optimization to determine the optimal training configuration for the proposed architecture. As shown in \autoref{tab_hp}, we considered four optimizers: Adam, AdamW, SGD, and RMSprop, a continuous learning rate boundary between 0.0001 and 0.01, and batch sizes ranging from 32 to 128. We found Adam to be the best optimizer with a learning rate of 0.005 and a batch size of 32 at the end of the experiment.

\subsection{TinyML Model Preparation}

For the TinyML, we converted the standard model into a TensorFlow Lite (TFLite) model that further reduces the computational load to test the model on resource-constrained devices. During conversion, the sample data was cast to 32-bit floating point numbers. A representative dataset compatible with the lite model was built using these samples with a batch size of 1. The model itself was built with the default post-training quantization option. To validate our network's capability, we deployed our TFLite model on a Raspberry Pi 4B and an Android OS 12 device powered by Exynos 9611 processor and evaluated its performance in real time.

\section{Results and Discussion}

\begin{table}[]
\centering
\caption{Performance comparison of different conventional models' performance compared to ours}
\label{tab_perf_comp}
\begin{tabular}{@{}lllll@{}}
\toprule
Model & Accuracy & Sensitivity & Specificity & \#Params \\ \midrule
YAMNet \citep{maity_enhancing_2024} & 0.57 & 0.30 & 0.82 & 3.75 M \\
AlexNet \citep{manshadi_murmur_2024} & 0.93 & 0.91 & -- & 60 M \\
Care4MyHeart \citep{alkhodari_ensemble_2022} & 0.76 & -- & -- & $\sim$65 M \\
SVM \citep{vimalajeewa_multiscale_2025} & 0.77 & 0.82 & 0.54 & -- \\
Revenger \citep{wen_searching_2022} & 0.74 & -- & -- & $\sim$19.8 M \\
VGG \citep{shin_temporal_2025} & 0.78 & 0.73 & 0.84 & 138 M \\
ConvNeXt \citep{shin_temporal_2025} & 0.83 & 0.74 & 0.91 & -- \\
MetaHeart \citep{xia_heart_2022} & 0.72 & -- & -- & -- \\
InceptionNet \citep{shin_temporal_2025} & 0.85 & 0.78 & 0.92 & $\sim$4 M \\
1D-CNN \citep{patwa_heart_2025} & 0.84 & 0.79 & 0.90 & $\sim$1 M \\
TCN \citep{shin_temporal_2025} & 0.88 & 0.81 & 0.95 & $\sim$70 k \\
IIITH \citep{venkataramani_modified_2022} & 0.71 & -- & -- & $\sim$25 M \\
LCNN \citep{han_deep_2024} & 0.80 & -- & -- & $\sim$5.5 M \\
Transformer \citep{alkhodari_identification_2024} & 0.90 & 0.72 & 0.88 & $\sim$65 M \\
ResMax \citep{han_deep_2024} & 0.77 & -- & -- & $\sim$300 k \\ \midrule
FunnelNet (Ours) & 0.85 & 0.85 & 0.92 & $\sim$5.4 k \\ \bottomrule
\end{tabular}
\end{table}

\subsection{Performance Comparison}

We evaluated the performance of our proposed FunnelNet against several existing models on the CirCor dataset. \autoref{tab_perf_comp} presents the accuracy, sensitivity, specificity, and total parameter counts for this comparison. 

Larger models with higher number of parameters achieved the highest overall metrics. For instance, \citet{manshadi_murmur_2024} obtained an accuracy of 0.93 and a sensitivity of 0.91 using AlexNet, while \citet{alkhodari_identification_2024} achieved an accuracy of 0.90 using an attention transformer-based approach. However, both models require approximately 60 million and 65 million parameters, respectively. Among lighter architectures, \citet{shin_temporal_2025} achieved an accuracy of 0.88 and a specificity of 0.95 using a temporal convolution (TCN) model with around 70k parameters. \citet{patwa_heart_2025} reported an accuracy of 0.84 using a 1D-CNN model with roughly 1 million parameters. Other approaches, such as the multi-task learning-based Revenger model by \citet{wen_searching_2022} and the transfer learning YAMNet by \citet{maity_enhancing_2024}, yielded lower accuracies of 0.74 and 0.57.

Our FunnelNet achieved an accuracy of 0.85, a sensitivity of 0.85, and a specificity of 0.92. While our accuracy is lower than the heavily parameterized AlexNet and Transformer models, it outperforms several other models, including the 1D-CNN, LCNN, and ConvNeXt architectures. Notably, we achieved these competitive results using only $\sim$5.4k parameters. This parameter count is substantially lower than all compared models, making our architecture highly efficient and practical for real-time deployment on resource-constrained edge devices.

\begin{figure*}[!tb]
    \centering
    \includegraphics[width=0.75\textwidth]{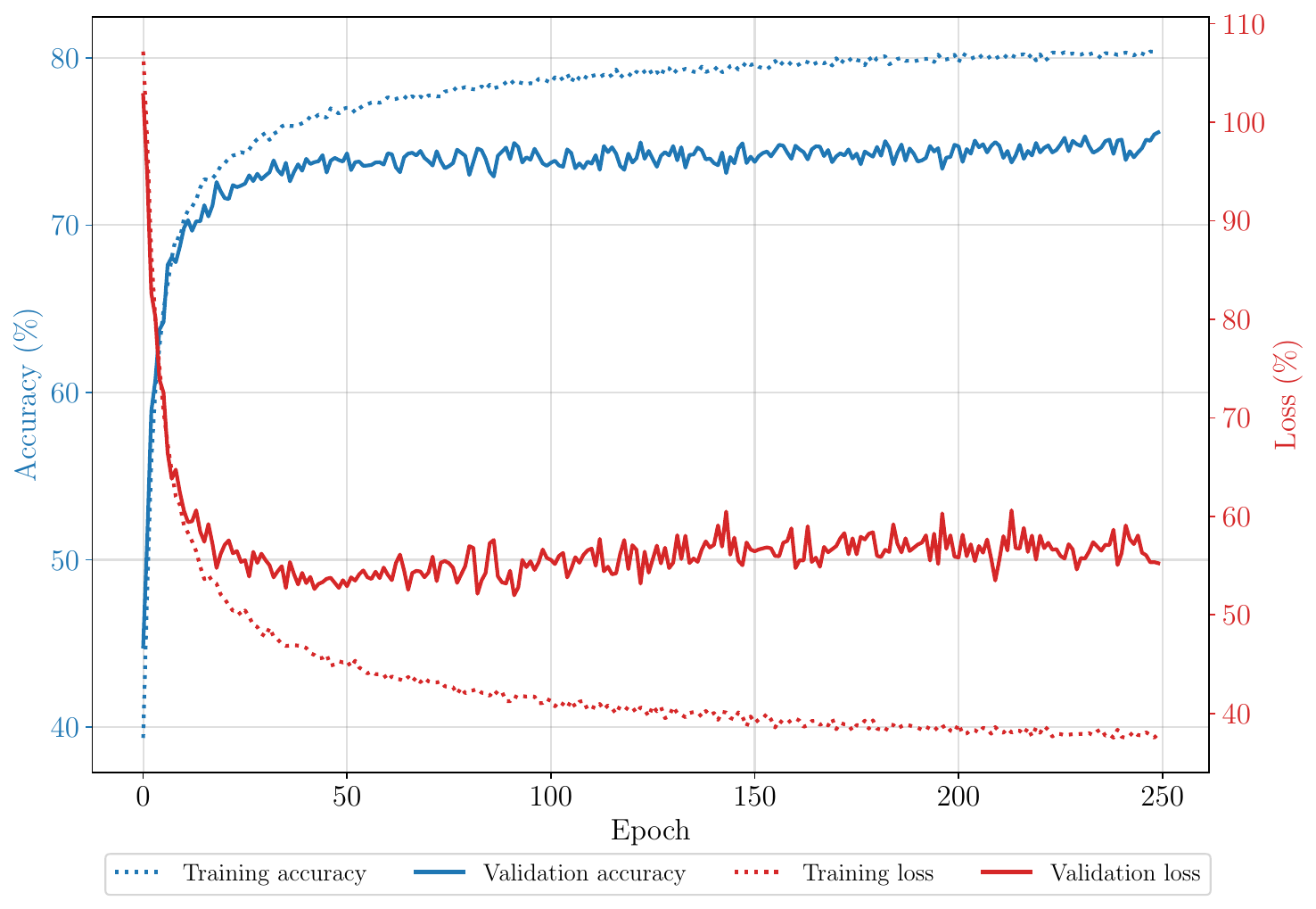}
    \caption{Average training and validation performance on the CirCor dataset over 250 epochs across all 10 folds of the cross-validation.}
    \label{fig_train_val_curve}
\end{figure*}

\begin{table}[!htb]
\centering
\caption{Ablation study on the model's performance for different configurations.}
\label{tab_ablation_study}
\begin{tabular}{@{}cclll@{}}
\toprule
\multicolumn{1}{l}{Outlier Removed} & \multicolumn{1}{l}{SMOTE} & Accuracy & Sensitivity & Specificity \\ \midrule
\xmark & \xmark & 0.70 & 0.37 & 0.70 \\
\cmark & \xmark & 0.68 & 0.34 & 0.67 \\
\xmark & \cmark & 0.33 & 0.33 & 0.67 \\
\cmark & \cmark & 0.85 & 0.85 & 0.92 \\ \bottomrule
\end{tabular}
\end{table}

\subsection{Ablation Study}

We conducted an ablation study to evaluate the individual and combined effects of outlier removal and SMOTE on the model's performance. \autoref{tab_ablation_study} summarizes the results across four configurations. The baseline approach, which used neither outlier removal nor SMOTE, achieved an accuracy of 0.70 and a low sensitivity of 0.37. Applying only outlier removal slightly decreased the accuracy to 0.68. Conversely, applying SMOTE without removing outliers caused a severe performance drop, reducing both accuracy and sensitivity to 0.33. This indicates that generating synthetic data from noisy, outlier-ridden samples actively harms the model's learning process. Finally, combining both outlier removal and SMOTE significantly improved all metrics, yielding an accuracy of 0.85, a sensitivity of 0.85, and a specificity of 0.92. The ablation results confirm that both the preprocessing steps are interdependent and essential for the proposed architecture to achieve optimal performance.

\subsubsection{Real-time Inference on Edge Device}


\begin{table}[]
\centering
\caption{Performance comparison of the lite version of our proposed model on two different devices.}
\label{tab_perf_comp_with_small_mdl}
\begin{tabular}{@{}lllll@{}}
\toprule
Device & Avg. Acc. & Avg. Inference Time & Min Inference Time & Max Inference Time \\ \midrule
Raspberry PI 4B & 0.91 & 9.3 ms & 9.1 ms & 16.3 ms \\
Android & 0.80 & 174 ms & 30 ms & 401 ms \\ \bottomrule
\end{tabular}
\end{table}

To demonstrate the practical applicability of our model, we deployed it on two edge devices: a Raspberry Pi 4B and an Android smartphone. We first converted our trained model into the TensorFlow Lite (TFLite) format to ensure efficient real-time inference on these resource-constrained platforms. \autoref{tab_perf_comp_with_small_mdl} details the performance of the lite model on both devices. On the Raspberry Pi 4B, we achieved an average accuracy of 0.91 with an average inference time of 9.3 ms, ranging from 9.1 ms to 16.3 ms. On the Android device, the model recorded an average accuracy of 0.80 and an average inference time of 174 ms.

Notably, the TinyML model on the Raspberry Pi achieved a slightly higher average accuracy compared to the standard model's performance. This minor performance variation likely results from the TFLite conversion process and internal optimization operations, such as post-training quantization which inadvertently acted as a regularizer to reduce noise and marginally improve the model's predictive accuracy.

\section{Conclusion}
\label{sec:conclusion}

In conclusion, our research presents an efficient and lightweight deep learning framework for real-time heart murmur detection. We processed the audio data using a Butterworth filter and the Morlet transform to eliminate noise and extract robust features. Our proposed FunnelNet, utilizing traditional and depthwise separable CNNs, showed comparable performance on the CirCor dataset while maintaining an extremely low parameter profile. Using only $\sim$5.4k parameters, our model achieved an accuracy of 0.85, a sensitivity of 0.85, and a specificity of 0.92, outperforming several heavier state-of-the-art architectures. Furthermore, we proved the model's practical viability for low-resource healthcare settings by converting it into a TinyML model for edge deployment. Real-time evaluation yielded an average accuracy of 0.91 on a Raspberry Pi 4B and 0.80 on an Android device. While our model balances predictive performance with minimal computational overhead, there remains room for future improvement. Because our current evaluation focused on a single dataset, future work should involve incorporating a more diverse range of heart murmur recordings to enhance the model's generalization capabilities and robustness. Additionally, exploring advanced signal processing techniques beyond the Morlet CWT and Butterworth filter could further optimize feature extraction and improve the model's diagnostic accuracy in real-world scenarios. As medical grade edge devices highly rely on distributed architectures, future work will also explore the secure and post-quantum-resilient frameworks (\citet{kamal2026quantum}) to support trustworthy and scalable real-world deployment by ensuring data security and integrity. 

\section*{Acknowledgement}

The authors would like to thank Helge Johan Risa from Linköping University for providing the embedded hardware resources used for model inference and performance testing.

\section*{CRediT Authorship Contribution Statement}

\textbf{Md Jobayer:} Writing -- original draft, Writing -- review and editing, Software, Methodology, Visualization, Validation, Conceptualization. \textbf{Md Mehedi Hasan Shawon:} Data curation, Writing -- original draft, Formal analysis. \textbf{Md Zakir Hossain:} Conceptualization, Writing -- review and editing, Supervision. \textbf{Shreya Ghosh:} Conceptualization, Writing -- review and editing, Data curation. \textbf{Imre Rudas:} Writing -- review and editing, Supervision. \textbf{Tom Gedeon:} Writing -- review and editing, Supervision. \textbf{Md Rakibul Hasan:} Conceptualization, Writing -- original draft, Writing -- review and editing, Software, Visualization, Supervision, Project administration.

\section*{Ethics Approval}
Not applicable

\section*{Funding Sources}

This research did not receive any specific grant from funding agencies in the public, commercial, or not-for-profit sectors.

\section*{Declaration of Competing Interest}

The authors declare that they have no known competing financial interests or personal relationships that could have appeared to influence the work reported in this paper.

\bibliographystyle{elsarticle-num-names}
\bibliography{ref.bib}

\end{document}